\documentclass[12pt,a4paper]{article}
\usepackage{amsmath}
\usepackage{cite}
\usepackage{graphicx}
\usepackage{amsfonts}
\usepackage{amssymb}
\newtheorem{montheo}{Rule}
\newtheorem{theo}{Claim}
\def\b{b^{-2}}

\def\my6j#1#2#3#4#5#6{
\left\{
\begin{array}
[c]{cc}%
#1 & #3\\
#2 & #4%
\end{array}
\mid
\begin{array}{c}
#5\\
#6%
\end{array}\right\}_{b}}

\def\fus#1#2#3#4#5#6{
F_{#5#6}^{L}\left[
\begin{array}
[c]{cc}%
#3 & #2\\
#4 & #1%
\end{array}
\right]}

\def\mon#1#2#3#4#5#6{
M^{-\frac{1}{2b}}_{#5#6}\left[
\begin{array}
[c]{cc}%
#3 & #2\\
#4 & #1%
\end{array}
\right]}

\def\fush#1#2#3#4#5#6{
F^{H_3^{+}}_{#5#6}\left[
\begin{array}
[c]{cc}%
#3 & #2\\
#4 & #1%
\end{array}
\right]}

\def\fussu2#1#2#3#4#5#6{
F^{SU(2)}_{#5#6}\left[
\begin{array}
[c]{cc}%
#3 & #2\\
#4 & #1%
\end{array}
\right]}

\def\braid#1#2#3#4#5#6{
B^{-\epsilon}_{#5#6}\left[
\begin{array}
[c]{cc}%
#3 & #2\\
#4 & #1%
\end{array}
\right]}

\def\sl2{
\hat{sl}(2)}

\def\al{
\alpha }

\def\de{
\Delta^{L} }

\def\Ga{
\Gamma_{b} }

\def\up{\Upsilon_{b}}

\def\uq{\mathcal{U}_q(\mathfrak{sl}(2,\mathbb{R}))}

\begin{document}
\bigskip
\hfill\hbox{AEI 2002-029} \vspace{2cm}

\begin{center}
{\Large \textbf{Monodromy of solutions of the}} \\
\vspace{0.3cm} {\Large\textbf{Knizhnik-Zamolodchikov equation:}}\\
 \vspace{0.3cm} {\Large\textbf{$SL(2,\mathbb{C})/SU(2)$
WZNW model}}

 \vspace{1cm}
{\Large B\'en\'edicte Ponsot\footnote{{\it e-mail address:
bponsot@aei-potsdam.mpg.de}}} \vspace{0.8cm}

{\it Max-Planck-Institut f\"ur Gravitationsphysik, Albert Einstein
Institut,

Am M\"uhlenberg 1,

D-14476 Golm, Germany.}

\vspace{2cm}
\end{center}

\begin{abstract}
Three explicit and equivalent representations for the monodromy of
the conformal blocks in the $SL(2,\mathbb{C})/SU(2)$ WZNW model
are proposed in terms of the same quantity computed in Liouville
field theory. We show that there are two possible fusion matrices
in this model. This is due to the fact that the conformal blocks,
being solutions to the Knizhnik-Zamolodchikov equation, have a
singularity when the $SL(2,\mathbb{C})$ isospin coordinate $x$
equals the worldsheet variable $z$. We study the asymptotic
behaviour of the conformal block  when $x$ goes to $z$. 
The obtained relation inserted into a four point correlation function in
the $SL(2,\mathbb{C})/SU(2)$ WZNW model gives some expression in terms
of two correlation functions in Liouville field theory.
\end{abstract}

\section*{Introduction}
The $SL(2,\mathbb{C})/SU(2)$ (or $H_3^+$) WZNW model is the second
simplest non compact Conformal Field Theory ({\it i.e.} with a
continuous spectrum of primary fields) besides Liouville field
theory. This model plays a role in condensed matter physics, as it
is believed to describe the plateau transitions
 in the Integer Quantum Hall effect \cite{BKSTT}. It is also intensively studied in the context of string
 theory on $AdS_3$ (see for example \cite{PST} for a non exhaustive list of references). As its characteristics
 are closely related to those of Liouville field theory,
one may wonder to what extent it is possible to apply to the $H_3^+$ WZNW model 
the techniques developped
for Liouville field theory (computation of the bulk three point function:\cite{DO,AAl}, bulk one point function
in presence of a boundary and boundary two point function \cite{FZZ},
Liouville field theory on the pseudosphere \cite{ZAZA}, bulk-boundary function \cite{HO}, boundary three point
 function \cite{PT3}, fusion coefficients and proof for crossing symmetry 
 \cite{PT1,PT2}).
  It turns out in the former model that the construction for the structure constants \cite{T2,PST}
can be achieved {\it mutatis mutandis} along the same lines. Less
staightforward is a proof for crossing symmetry,
 as well as the construction of the fusion matrix. In Liouville field theory, the fusion matrix was constructed as
  Racah-Wigner coefficients for some well chosen continuous representations 
  of the quantum group $\uq$ \cite{PT1,PT2}; in this case the proof
  for crossing symmetry boiled down to an orthogonality relation  for these 
  Racah-Wigner coefficients. In the
  $H_3^+$ WZNW model, it was proposed in \cite{FM} as a quantum group structure the "pair" 
  $U_q(sl(2)),U_{q'}(osp(1|2))$. We will not follow this approach here,
   as it turns out that
   there is a way to avoid quantum group methods to construct the fusion matrix:
    as it was noticed in \cite{BKSTT} and \cite{T1}, it is possible to
    adapt to the $H_3^+$ WZNW model some observation previously made by Zamolodchikov and Fateev in \cite{FZ} in the
     context of the compact $SU(2)$ WZNW model: these authors observed that there exists a 5 point conformal block
      in the
     (k+2,1) Minimal Model that satisfies the same Knizhnik-Zamolodchikov equation \cite{KZ} as the 4 point conformal block
     in the $SU(2)$ WZNW model. The adaptation of this relation to the non compact case is straighforward and
     allowed to prove
 crossing symmetry in the $H_3^{+}$ WZNW model \cite{T1}, as a consequence from a similar property of a 5 point function
 in Liouville field theory. Actually, this method also permits\footnote{This was first observed by J.~Teschner.}
 to construct the monodromy of the conformal blocks in the $H_3^+$ WZNW model 
 ({\it i.e.} the fusion matrix, also called fusion coefficients) from the
  knowledge of the monodromy for a special 5 point function in Liouville field theory. Let us emphasize on the fact
   that the fusion matrix is the most important quantity of a  CFT on genus zero, as it encodes the information
   on degenerate representations, fusion rules, intermediate states appearing in a four point function; it also
    permits the computation of boundary structure constants and enters the consistency relations satisfied by
    structure constants \cite{BPPZ,R}.\\
In the work\footnote{I thank V.~B.~Petkova for pointing out this reference, as well as the work \cite{PRY}.} \cite{FGP} is covered the case where one of the external spins corresponds to a degenerate $\sl2$ representation\footnote{Degeneracy arises due to the existence of null vectors.}, the other spins being generic. The fusion matrix is derived explicitly thanks to the explicit construction of the conformal blocks in this case (see also \cite{PRY}),
enlarging the method of screening integrals developped in \cite{DF}. These conformal blocks reproduced the fusion rules for $\sl2$ given in \cite{AY}.\\
In the present article, we will consider the problem in its full generality when all the external spins are generic: there is no formula known for the conformal blocks in general and the monodromy is infinite dimensional. The paper is organized as follows: in the first section we will
recall some useful information about Liouville field theory and
the $H_3^+$ WZNW model. In section two we will propose three
explicit formulas for the monodromy of the conformal blocks in
$H_3^+$ WZNW, constructed in terms of two Liouville fusion matrices.
Let us mention that the case we have here is
 somewhat different to all other known cases so far ({\it i.e.} minimal models, $SU(2)$ WZNW model, Liouville field theory), as there are
 two possible fusion matrices for this model\footnote{We consider here the two most natural choices for the cuts, as in \cite{MS}.} (this was first noticed in \cite{FGP,PRY}): this comes from the fact that the conformal blocks, in addition to the
  usual singularities at $z=0,1,\infty$, have also a singularity at $z=x$ (this plays an important role in \cite{MO}),
  where $z$ is the worldsheet variable, and $x$ the isospin variable (or space-time coordinate). When deriving the
  formula for the fusion matrix, there are two cases to consider, 
  depending whether $Im(z-x)$ is positive or negative.
   This accounts for a phase factor in the formulas, which is different
    according
   to the sign of
   $Im(z-x)$.
    In section three, we will study the asymptotic behavior of the conformal
 blocks in the limit where $x$ goes to $z$, and we will see that they can be expressed as a sum of two Liouville
  conformal blocks. The first term is regular in this limit, whereas the second term contains some singularity. This limit is interpreted as quantum hamiltonian reduction in \cite{FGPP} (with the external spins satisfying charge
conservation conditions).
  The obtained relation permits us to rewrite a four point correlation function in the limit $x$ goes to $z$ as a sum over
  two Liouville correlation functions. We finish by some concluding remarks; appendix A contains some particular
   cases of the Liouville fusion coefficients needed in the main text to derive our main result, in appendix B we
   recover some well known special cases from our $H_3^+$ WZNW fusion matrix. Finally, in appendix C we present a way
   to find degenerate representations and fusion rules for $\sl2$ from degenerate representations and fusion rules
   for the Virasoro algebra.

\section{Requisites}
\subsection{Liouville field theory}
Let the $V_{\al}(z,\bar{z})$ be the primary fields
 with conformal weight $\de_{\al}=\al(Q-\al)$ where $Q=b+b^{-1}$; $b$ 
 is the coupling constant in Liouville field theory that we shall call for short 
 LFT. The central charge of the Virasoro algebra is $c=1+6Q^{2}$.\\
Let
\begin{eqnarray}
\left\langle V_{\al_4}(\infty)V_{\al_3}(1)V_{\al_{2}}(z,\bar{z})V_{\al_1}(0)\right\rangle
\equiv \mathcal{V}_{\al_4,\al_3,\al_2,\al_1}(z,\bar{z}) \nonumber
\end{eqnarray}
denote a four point correlation function in LFT and
\begin{eqnarray}
\mathcal{F}^{s}_{\al_{21}}(\al_1,\al_2,\al_3,\al_4|z)
\end{eqnarray}
be the corresponding conformal block in the s-channel. 
The conformal block is completly determined by the conformal symmetry (although there
is no known closed form for it in general),
 and is normalized such that
\begin{eqnarray}
\mathcal{F}^{s}_{\al_{21}}(\al_1,\al_2,\al_3,\al_4|z)\sim_{z\rightarrow 0}z^{\de_{\al_{21}}-\de_{\al_1}-\de_{\al_2}}(1+\mathcal{O}(z))
\end{eqnarray}
Let us note that the Liouville conformal block depends on conformal weights only.\\
There exist \cite{T} invertible fusion transformations between s-
and t-channel conformal blocks, defining the Liouville fusion matrix:
\begin{eqnarray}
\mathcal{F}^{s}_{\al_{21}}(\al_1,\al_2,\al_3,\al_4|z)=\int_{\frac{Q}{2}+i\mathbb{R}^{+}}d\al_{32}\fus{\al_1}{\al_2}{\al_3}{\al_4}{\al_{21}}{\al_{32}}\mathcal{F}^{t}_{\al_{32}}(\al_1,\al_2,\al_3,\al_4|1-z)
\label{transfost}
\end{eqnarray}
The explicit expression for the LFT fusion matrix was given in \cite{PT1}, in terms of a $b$-deformed $_{4}F_{3}$ hypergeometric function in the Barnes representation:
\begin{eqnarray}
\lefteqn{\fus{\al_1}{\al_2}{\al_3}{\al_4}{\al_{21}}{\al_{32}}=} \nonumber \\
&&
\frac{\Ga(2Q-\al_3-\al_2-\al_{32})\Ga(\al_3+\al_{32}-\al_2)\Ga(Q-\al_2-\al_{32}+\al_3)\Ga(Q-\al_3-\al_2+\al_{32})}{\Ga(2Q-\al_1-\al_2-\al_{21})\Ga(\al_1+\al_{21}-\al_2)\Ga(Q-\al_2-\al_{21}+\al_1)\Ga(Q-\al_2-\al_1+\al_{21})} \nonumber \\
&&
\frac{\Ga(Q-\al_{32}-\al_1+\al_4)\Ga(\al_{32}+\al_{1}+\al_{4}-Q)\Ga(\al_1+\al_4-\al_{32})\Ga(\al_4+\al_{32}-\al_1)}{\Ga(Q-\al_{21}-\al_{3}+\al_4)\Ga(\al_{21}+\al_3+\al_4-Q)\Ga(\al_3+\al_4-\al_{21})\Ga(\al_{21}+\al_4-\al_3)} \nonumber \\
&&\frac{\Ga(2Q-2\al_{21})\Ga(2\al_{21})}{\Ga(Q-2\al_{32})\Ga(2\al_{32}-Q)}
\frac{1}{i}\int\limits_{-i\infty}^{i\infty}ds \;\;
\frac{S_b(U_1+s)S_b(U_2+s)S_b(U_3+s)S_b(U_4+s)}
{S_b(V_1+s)S_b(V_2+s)S_b(V_3+s)S_b(Q+s)} \nonumber \\
\label{fusionmatrix}
\end{eqnarray}
with
$$
\begin{array}{ll}
 U_1 = \al_{21}+\al_1-\al_2                &  V_1 = Q+\al_{21}-\al_{32}-\al_{2}+\al_{4} \\
 U_2 = Q+\al_{21}-\al_2-\al_1              &  V_2 = \al_{21}+\al_{32}+\al_{4}-\al_2 \\
 U_3 = \al_{21}+\al_{3}+\al_{4}-Q          &  V_3 = 2\al_{21} \\
 U_4 = \al_{21}-\al_{3}+\al_4 \\              \\
\end{array}
$$
The special function entering the formula above is
$\Gamma_b(x)\equiv
\frac{\Gamma_2(x|b,b^{-1})}{\Gamma_2(Q/2|b,b^{-1})}$, where
$\Gamma_2(x|\omega_1,\omega_2)$ is the Double Gamma function introduced by Barnes
\cite{Barnes}, which definition is
\begin{eqnarray}
\nonumber \\
&& \text{log}\Gamma_{2}(x|\omega_1,\omega_2)=\left(\frac{\partial}{\partial t}\sum_{n_1,n_2=0}^{\infty}(x+n_1\omega_1+n_2\omega_2)^{-t}\right)_{t=0} \nonumber
\end{eqnarray}
The function $\Ga(x)$ such defined is such that $\Gamma_b(x)\equiv\Gamma_{b^{-1}}(x)$, and satisfies the following functional relation
\begin{eqnarray}
\Ga(x+b)= \frac{\sqrt{2\pi}b^{bx-\frac{1}{2}}}{\Gamma(bx)}\Ga(x)
\end{eqnarray}
$\Ga(x)$ is a meromorphic function of $x$, which poles are located at
$x=-nb-mb^{-1}, n,m \in \mathbb{N}$. The function $S_b(x)$ is defined as $S_b(x)\equiv \frac{\Ga(x)}{\Ga(Q-x)}$.\\
Let us quote the following properties:
\begin{itemize}
\item[-]
The Liouville fusion matrix is holomorphic in the following range \cite{PT1}
$$
\begin{array}{ll}
|\text{Re}(\al_2+\al_3+\al_{32}-Q)|<Q   &|\text{Re}(\al_4+\al_1+\al_{32}-Q)|<Q \\
|\text{Re}(\al_2+\al_3-\al_{32})|<Q     &|\text{Re}(\al_4+\al_1-\al_{32})|<Q \\
|\text{Re}(\al_3+\al_{32}-\al_{2})|<Q   &|\text{Re}(\al_4+\al_{32}-\al_{1})|<Q \\
|\text{Re}(\al_2+\al_{32}-\al_{3})|<Q     &|\text{Re}(\al_1+\al_{32}-\al_{4})|<Q \\
\end{array}
$$
\item[-]
It satisfies the symmetry properties
\begin{eqnarray}
\fus{\al_{1}}{\al_2}{\al_3}{\al_4}{\al_{21},}{\al_{32}}=\fus{\al_{2}}{\al_1}{\al_4}{\al_3}{\al_{21},}{\al_{32}}=\fus{\al_{4}}{\al_3}{\al_2}{\al_1}{\al_{21},}{\al_{32}}
\end{eqnarray}
\item[-]
As the conformal blocks depends on conformal weights only, so does
the Liouville fusion matrix, {\it i.e.} is invariant when one of
the $\al_i$ is substituted by $Q-\al_i$.
\end{itemize}
The LFT fusion matrix was built in terms of the Racah-Wigner coefficients for an appropriate continuous series of representations of the quantum group $\uq$ with deformation parameter $q=e^{i\pi b^{2}}$ \cite{PT1,PT2}. This construction {\it ensures} that the fusion matrix satisfies the set of Moore-Seiberg equations (or polynomial equations) \cite{MS}.\\
Let us quote the {\it pentagonal} equation:
\begin{align}
\int_{\frac{Q}{2}+i\mathbb{R^{+}}}d\delta_{1}\fus{\al_1}{\al_2}{\al_3}{\beta_{2}}{\beta_{1}}{\delta_{1}}
\fus{\al_1}{\delta_1}{\al_4}{\al_5}{\beta_{2}}{\gamma_{2}}
\fus{\al_2}{\al_3}{\al_4}{\gamma_{2}}{\delta_{1}}{\gamma_{1}}\nonumber \\
=\fus{\beta_1}{\al_3}{\al_4}{\al_5}{\beta_{2}}{\gamma_{1}}
\fus{\al_1}{\al_2}{\gamma_1}{\al_{5}}{\beta_{1}}{\gamma_{2}}
\label{pentagone}
\end{align}
and the {\it two hexagonal} equations:
\begin{align}
\lefteqn{\fus{\al_1}{\al_2}{\al_4}{\al_3}{\al_{21},}{\beta}e^{i\pi\epsilon(\de_{\al_1}+\de_{\al_2}+\de_{\al_3}+\de_{\al_4}-\de_{\al_{21}}-\de_{\beta)}}=} \nonumber \\
&& \int_{\frac{Q}{2}+i\mathbb{R}^{+}}d\al_{32}\fus{\al_1}{\al_2}{\al_3}{\al_4}{\al_{21},}{\al_{32}}\fus{\al_1}{\al_4}{\al_2}{\al_3}{\al_{32},}{\beta}e^{i\pi \epsilon\Delta_{\al_{32}}}
\end{align}
where $\epsilon =\pm$.\\
We will also need the Liouville three point function; an explicit formula for it was proposed in
 \cite{DO,AAl}
\begin{eqnarray}
\lefteqn{C^{L}(\al_3,\al_2,\al_1)= \left\lbrack \pi \mu \gamma(b^{2})b^{2-2b^{2}}\right\rbrack^{\frac{Q-\al_1-\al_2-\al_3}{b}}}. \nonumber \\
&&  \frac{\Upsilon_0\up(2\al_1)\up(2\al_2)\up(2\al_3)}{\up(\al_1+\al_2+\al_3-Q)\up(\al_1+\al_2-\al_3)\up(\al_1+\al_3-\al_2)\up(\al_2+\al_3-\al_1)} \nonumber \\
\label{3pointsZZ}
\end{eqnarray}
where $\up(x)^{-1}=\Gamma_b(x)\Gamma_b(Q-x), \gamma(x)=\frac{\Gamma(x)}{\Gamma(1-x)}, \Upsilon_{0}=\text{res}_{x=0}\frac{d\up(x)}{dx}$, and $\mu$ is the cosmological constant.\\
A four point function in LFT is written (if the $Re(\al_{i}), i=1\dots 4$ are close enough to $Q/2$)
\begin{eqnarray}
\lefteqn{\mathcal{V}_{\al_4,\al_3,\al_2,\al_1}(z,\bar{z}) =}\nonumber \\
&& =\frac{1}{2}\int_{\frac{Q}{2}+i\mathbb{R}}d\al_{21}C^{L}(\al_{4},\al_3,\al_{21})C^{L}(Q-\al_{21},\al_2,\al_1)|\mathcal{F}^{s}_{\al_{21}}(\al_1,\al_2,\al_3,\al_4|z)|^{2}
\end{eqnarray}

\subsection{$SL(2,\mathbb{C})/SU(2)$ WZNW model}
Let us denote $k$ the level of the current algebra; $k$ is formally related to the coupling constant $b$ in LFT by
 $k\equiv b^{-2}+2$. The central charge of the theory is $c=\frac{3k}{k-2}$. The primary fields $\Phi^{j}(x|z)$ have conformal weight $\Delta_{j}=-b^{2}j(j+1)$.\\
The action of the $SL(2,\mathbb{C})$ currents on the primary fields is given by
\begin{eqnarray}
J^{a}(z)\Phi^{j}(x|w)\sim\frac{D^{a}_{j}}{z-w}\Phi^{j}(x|w), \quad a=\pm,3 \qquad \bar{J}^{a}(z)\Phi^{j}(x|w)\sim\frac{\bar{D}^{a}_{j}}{\bar{z}-\bar{w}}\Phi^{j}(x|w)
\end{eqnarray}
where $D^{a}_{j}$ are differential operators representing the $sl(2)$ algebra
\begin{eqnarray}
D^{+}_{j}=\frac{\partial}{\partial x}, \qquad D^{3}_{j}=x\frac{\partial}{\partial x}+j,\qquad  D^{-}_{j}=x^{2}\frac{\partial}{\partial x}+2jx,
\end{eqnarray}
the $\bar{D}^{a}_{j}$ their complex conjugates.\\
Let
\begin{eqnarray}
\left\langle \Phi^{j_4}(\infty|\infty) \Phi^{j_3}(1|1)\Phi^{j_2}(x|z)\Phi^{j_1}(0|0)\right\rangle
\equiv \Phi_{j_4,j_3,j_2,j_1}(x,\bar{x}|z,\bar{z})\nonumber
\end{eqnarray}
be a four point correlation function in the
$SL(2,\mathbb{C})/SU(2)$ WZNW model and
\\
$\mathcal{G}^{s}_{j_{21}}(j_1,j_2,j_3,j_4|x,z)$ be the
corresponding s-channel conformal block. It is uniquely defined as
the solution of the Knizhnik-Zamolodchikov equation
$(z(z-1)\partial_z+b^{2}\mathcal{D}^{(2)}_x)\mathcal{G}^{s}(x|z)=0$,
where
\begin{eqnarray}
\lefteqn{\mathcal{D}^{(2)}_x=x(x-1)(x-z)\partial_x^{2}}\nonumber \\
&& -[(\kappa-1)(x^2-2zx+z)+2j_1x(z-1)+2j_2x(x-1)+2j_3z(x-1)]\partial_x\nonumber \\
&&+2j_2\kappa (x-z)+2j_1j_2(z-1)+2j_2j_3z
\end{eqnarray}
where $\kappa=j_1+j_2+j_3-j_4$, and the normalization prescription
\begin{eqnarray}
\mathcal{G}^{s}(x|z)\sim z^{\Delta_{j_{21}}-\Delta_{j_{2}}-\Delta_{j_{1}}}x^{j_1+j_2-j_{21}}(1+\mathcal{O}(x)+\mathcal{O}(z)).
\end{eqnarray}
in the limit of taking first $z\rightarrow 0$, then $x\rightarrow
0$. Let us note that the solutions of the KZ equation have four singular points, located
 at $z=0,1,x,\infty$, and that there is no closed form known for the conformal blocks 
 in general. \\ The
monodromy of the conformal blocks (or fusion matrix of the
$H_3^{+} $ WZNW model) is defined as
\begin{eqnarray}
\mathcal{G}^{s}_{j_{21}}(j_1,j_2,j_3,j_4|x,z)=\int_{-\frac{1}{2}-i\infty}^{-\frac{1}{2}+i\infty}dj_{32}\fush{j_1}{j_2}{j_3}{j_4}{j_{21}}{j_{32}}\mathcal{G}^{t}_{j_{32}}(j_1,j_2,j_3,j_4|1-x,1-z)
\end{eqnarray}
It is our aim to compute this quantity.\\
We will also need the expression for the three point function \cite{T2}
\begin{eqnarray}
\lefteqn{C^{H_3^{+}}(j_3,j_2,j_1)= (\nu
(b)b^{-2b^{2}})^{1+j_1+j_2+j_3}}\nonumber \\
&& \frac{C_0(b)
\up(-2bj_1)\up(-2bj_2)\up(-2bj_3)}{\up(-bj_1-bj_2-bj_3-b)\up(-bj_1-bj_2+bj_3)\up(-bj_1-bj_3+bj_2)\up(-bj_2-bj_3+bj_1)} \nonumber \\
\label{3pointsH}
\end{eqnarray}
It is known \cite{T2} that the expression for a four point
correlation function is (if the $Re(j_{i}), i=1\dots 4$ are close
enough to $-\frac{1}{2}$).
\begin{eqnarray}
\lefteqn{\Phi_{j_4,j_3,j_2,j_1}(x,\bar{x}|z,\bar{z}) =}\nonumber \\
&& \int_{-\frac{1}{2}+i\mathbb{R}}dj_{21}B(j_{21})C^{H_3^{+}}(j_4,j_3,j_{21})C^{H_3^{+}}(j_{21},j_2,j_1)|
\mathcal{G}^{s}_{j_{21}}(j_1,j_2,j_3,j_4|x,z)|^{2}
\end{eqnarray}
where
$$B(j)=(\nu(b))^{-2j-1}\gamma(1+b^{2}(2j+1)).$$

\section{Monodromy of solutions of the KZ equation}

\subsection{Method}
It follows from a remarkable observation of \cite{FZ} straightforwardly adapted to the non-compact case
 that the conformal block of a special 5 point function in LFT satisfies {\it the same KZ equation} as
 the conformal block in the $H_3^{+}$ WZNW model (for an explanation of this relation, see \cite{S});
  hence, knowing the precise correspondence between the two quantities allows us to compute the monodromy of
  the conformal block in the $H_3^{+}$ WZNW model in terms of the monodromy of this 5 point LFT conformal block.\\
Let $\mathcal{F}^{s}_{\al_{21}}(\al_1,\al_2,-\frac{1}{2b},\al_3,\al_4|x,z)$
be the 5 point conformal block corresponding to the LFT correlation function:
\begin{eqnarray}
\left\langle V_{\al_4}(\infty)V_{\al_3}(1)V_{-\frac{1}{2b}}(x,\bar{x})V_{\al_{2}}(z,\bar{z})V_{\al_1}(0)\right\rangle \nonumber
\end{eqnarray}
it was shown in \cite{FZ} that
\begin{eqnarray}
\lefteqn{\mathcal{F}^{s}_{\al_{21}}(\al_1,\al_2,-\frac{1}{2b},\al_3,\al_4|x,z)=}\nonumber \\
&&x^{b^{-1}\al_1}(1-x)^{b^{-1}\al_3}(x-z)^{b^{-1}\al_2}z^{\frac{1}{2}\gamma_{12}}(1-z)^{\frac{1}{2}\gamma_{13}}\mathcal{G}^{s}_{j_{21}}(j_1,j_2,j_3,j_4|x,z)
\label{corres}
\end{eqnarray}
where the parameters of the two theories are related by
$$
\begin{array}{ll}
2\al_1=-b(j_1+j_2-j_3-j_4-b^{-2}-1)  &  2\al_3=-b(-j_1+j_2+j_3-j_4-b^{-2}-1)\\
2\al_2=-b(j_1+j_2+j_3+j_4+1)         &  2\al_4=-b(-j_1+j_2-j_3+j_4-b^{-2}-1)\\
\al_{21}=-bj_{21}+\frac{1}{2b}       &  \al_{32}=-bj_{32}+\frac{1}{2b}\\
\gamma_{12}=4b^{2}j_1j_2-4\al_1\al_2 &  \gamma_{23}=4b^{2}j_2j_3-4\al_2\al_3\\
\end{array}
\label{parametres}
$$
One reads off from (\ref{corres}) the following relation between the two monodromies:
\begin{eqnarray}
\fush{j_1}{j_2}{j_3}{j_4}{j_{21},}{j_{32}}_{\epsilon}=e^{i\pi\epsilon
b^{-1}\al_2}\mon{\al_1}{\al_2}{\al_3}{\al_4}{\al_{21},}{\al_{32}}_{\epsilon}
\end{eqnarray}
where $\epsilon=\pm1$ depends whether $Im(x-z)$ is negative or positive (this was first seen in \cite{FGP,PRY}).
 This leads to an ambiguity in the definition of the $H_3^{+}$ WZNW fusion matrix,
 for this reason we shall add to the fusion matrix the subscript $\epsilon$. Nevertheless, this ambiguity does not appear
  in the bootstrap, where one considers both the holomorphic conformal block and its antiholomorphic counterpart.

\subsection{Monodromy of the Liouville 5 point conformal block}
The monodromy of this special 5 point conformal block decomposes in a succession of elementary braiding and fusing
transformations.\\
Let us remember that the braiding is related to the fusion the following way \cite{MS}:
\begin{eqnarray}
\braid{\alpha_1}{\alpha_2}{\al_4}{\alpha_3}{\alpha_{21},}{\alpha_{32}}=e^{i\pi\epsilon(\Delta^{L}_{\al_{21}}+\Delta^{L}_{\al_{32}}-\Delta^{L}_{\al_{3}}-\Delta^{L}_{\al_4})}
\fus{\alpha_1}{\alpha_2}{\al_3}{\alpha_4}{\alpha_{21},}{\alpha_{32}}
\end{eqnarray}
It is then straightforward to obtain the monodromy for this 5 point conformal block thanks to
the picture:

\unitlength=0.8cm
\begin{picture}(22,4)
\put(0,1){\line(1,0){6.5}}
\put(1,1){\line(0,1){1.5}}
\put(3,1){\line(0,1){1.5}}
\put(5,1){\line(0,1){1.5}}
\put(0,1.2){$\alpha_{1}$}
\put(0.5,2.5){$\alpha_{2}$}
\put(2.5,2.7){$-\frac{1}{2b}$}
\put(5.2,2.5){$\alpha_{3}$}
\put(6.2,1.2){$\alpha_{4}$}
\put(1.2,1.2){$\alpha_{21}$}
\put(3.2,1.2){$\alpha_{21}-\frac{1}{2b}$}
\put(7.2,1.2){$=\sum_{s=\pm}\braid{\alpha_1}{\alpha_2}{-\frac{1}{2b}}{\alpha_{21}-\frac{1}{2b}}{\alpha_{21},}{\alpha_1-s\frac{1}{2b}}$}
\put(16,1){\line(1,0){6.5}}
\put(17,1){\line(0,1){1.5}}
\put(19,1){\line(0,1){1.5}}
\put(21,1){\line(0,1){1.5}}
\put(16,1.2){$\alpha_{1}$}
\put(16.5,2.7){$-\frac{1}{2b}$}
\put(18.5,2.5){$\alpha_{2}$}
\put(21.2,2.5){$\alpha_{3}$}
\put(22.2,1.2){$\alpha_{4}$}
\put(17.2,1.2){$\alpha_{1}-s\frac{1}{2b}$}
\put(19.2,1.2){$\alpha_{21}-\frac{1}{2b}$}
\end{picture}
\\
\begin{picture}(23,4)
\put(0,1){\line(1,0){6.5}}
\put(1,1){\line(0,1){1.5}}
\put(3,1){\line(0,1){1.5}}
\put(5,1){\line(0,1){1.5}}
\put(0,1.2){$\alpha_{1}$}
\put(0.5,2.7){$-\frac{1}{2b}$}
\put(2.5,2.5){$\alpha_{2}$}
\put(5.2,2.5){$\alpha_{3}$}
\put(6.2,1.2){$\alpha_{4}$}
\put(1.2,1.2){$\alpha_{1}\pm\frac{1}{2b}$}
\put(3.2,1.2){$\alpha_{21}-\frac{1}{2b}$}
\put(7.2,1.2){$=\int d\al_{32}\fus{\alpha_1-s\frac{1}{2b}}{\alpha_2}{\alpha_3}{\alpha_4}{\alpha_{21}-\frac{1}{2b},}{\alpha_{32}}$}
\put(16,1){\line(1,0){6.5}}
\put(17,1){\line(0,1){1.5}}
\put(20,1){\line(0,1){1}}
\put(20,2){\line(1,0){2.5}}
\put(20.5,2){\line(0,1){1}}
\put(16,1.2){$\alpha_{1}$}
\put(16.5,2.7){$-\frac{1}{2b}$}
\put(20,3){$\alpha_{2}$}
\put(22.2,2.2){$\alpha_{3}$}
\put(22.4,1.1){$\alpha_{4}$}
\put(17.2,1.2){$\alpha_{1}\pm\frac{1}{2b}$}
\put(20,2.2){$\alpha_{32}$}
\end{picture}
\\
\begin{picture}(22,4)
\put(0,1){\line(1,0){6.5}}
\put(1,1){\line(0,1){1.5}}
\put(4,1){\line(0,1){1}}
\put(4,2){\line(1,0){2.5}}
\put(4.5,2){\line(0,1){1}}
\put(0,1.2){$\alpha_{1}$}
\put(0.5,2.7){$-\frac{1}{2b}$}
\put(4,3){$\alpha_{2}$}
\put(6.2,2.2){$\alpha_{3}$}
\put(6.4,1.1){$\alpha_{4}$}
\put(1.2,1.2){$\alpha_{1}\pm\frac{1}{2b}$}
\put(4,2.2){$\alpha_{32}$}
\put(7.2,1.2){$=\fus{\alpha_1}{-\frac{1}{2b}}{\alpha_{32}}{\alpha_4}{\alpha_1-s\frac{1}{2b},}{\alpha_{32}-\frac{1}{2b}}$}
\put(16,1){\line(1,0){6.5}}
\put(17,1){\line(0,1){1}}
\put(17,2){\line(1,0){4.5}}
\put(18,2){\line(0,1){1.5}}
\put(20,2){\line(0,1){1.5}}
\put(16,1.2){$\alpha_{1}$}
\put(17.7,3.5){$-\frac{1}{2b}$}
\put(20.3,3.5){$\alpha_{2}$}
\put(21.5,2.2){$\alpha_{3}$}
\put(22.4,1.1){$\alpha_{4}$}
\put(17.2,1.2){$\alpha_{32}-\frac{1}{2b}$}
\put(19,2.3){$\alpha_{32}$}
\end{picture}
where the first picture (top left) represents the s-channel 5 point conformal block, and the last one (bottom right) the t-channel 5 point conformal block.\\
Collecting the terms together, we get:
\begin{eqnarray}
\lefteqn{\mon{\al_1}{\al_2}{\al_3}{\al_4}{\al_{21},}{\al_{32}}_{\epsilon}=}\nonumber \\
& & e^{i\pi\epsilon b^{-1}(-\al_{21}+\al_1)}
F^{L}_{-+}\left[
\begin{array}
[c]{cc}%
\al_{1}    & -\frac{1}{2b}\\
\al_{2}    &   \al_{21}-\frac{1}{2b}%
\end{array}
\right]
F^{L}_{++}\left[
\begin{array}
[c]{cc}%
\al_{32}   &  -\frac{1}{2b}\\
\al_{4}    &   \al_{1}%
\end{array}
\right]
F^{L}_{\al_{21}-\frac{1}{2b},\al_{32}}\left[
\begin{array}
[c]{cc}%
\al_{3}    &  \al_2 \\
\al_{4}    &   \al_{1} -\frac{1}{2b}%
\end{array}
\right] + \nonumber \\
& & e^{i\pi\epsilon b^{-1}(-\al_{21}-\al_1+Q)}
F^{L}_{--}\left[
\begin{array}
[c]{cc}%
\al_{1}    &  -\frac{1}{2b}\\
\al_{2}    &   \al_{21} -\frac{1}{2b}%
\end{array}
\right]
F^{L}_{-+}\left[
\begin{array}
[c]{cc}%
\al_{32}   &  -\frac{1}{2b}\\
\al_{4}    &   \al_{1}%
\end{array}
\right]
F^{L}_{\al_{21}-\frac{1}{2b},\al_{32}}\left[
\begin{array}
[c]{cc}%
\al_{3}    &  \al_2 \\
\al_{4}    &   \al_{1} +\frac{1}{2b}%
\end{array}
\right]\nonumber \\
\label{5point}
\end{eqnarray}

\subsection{Fusion matrix of the $H_3^+$ WZNW model}
The formula for the $H_3^+$ WZNW fusion matrix follows (see Appendix for this special value of the Liouville fusion coefficient when one of the $\al_i, i=1\dots 4$ is equal to $-b^{-1}/2$):
\begin{eqnarray}
\fush{j_1}{j_2}{j_3}{j_4}{j_{21},}{j_{32}}_{\epsilon}=e^{i\pi\epsilon
b^{-1}\al_2}\mon{\al_1}{\al_2}{\al_3}{\al_4}{\al_{21},}{\al_{32}}_{\epsilon}
\end{eqnarray}
\begin{eqnarray}
\lefteqn{\fush{j_1}{j_2}{j_3}{j_4}{j_{21},}{j_{32}}_{\epsilon}=} \nonumber \\
& & \frac{\pi \Gamma(2+b^{-2}+2j_{21})\Gamma(1+2j_{32})}{\sin\pi(j_1+j_2-j_3-j_4)}\times \nonumber \\
& &
\Big(\frac{e^{i\pi\epsilon(j_{21}-j_1-j_2)}}{\Gamma(2+b^{-2}+j_{21}+j_1+j_2)\Gamma(j_{21}-j_3-j_4)\Gamma(1+j_{32}-j_1+j_4)\Gamma(1+j_3-j_2+j_{32})}\times \nonumber \\
& & \times F^{L}_{-bj_{21},-bj_{32}+1/2b}\left[
\begin{array}
[c]{cc}%
-bj_3+1/2b    &  -bj_{2} \\
-bj_4+1/2b    &  -bj_{1}%
\end{array}
\right]
\nonumber \\
& & -\frac{e^{i\pi\epsilon(j_{21}-j_3-j_4)}}{\Gamma(2+b^{-2}+j_{21}+j_3+j_4)\Gamma(j_{21}-j_1-j_2)\Gamma(1+j_{32}+j_1-j_4)\Gamma(1-j_3+j_2+j_{32})} \times \nonumber \\
& & \times F^{L}_{-bj_{21},-bj_{32}+1/2b}\left[
\begin{array}
[c]{cc}%
-bj_{3}    &  -bj_{2}+1/2b \\
-bj_{4}    &  -bj_{1} +1/2b%
\end{array}
\right] \Big) \label{fush3+}
\end{eqnarray}
It is possible to obtain an alternative representation for the $H_3^+$ fusion matrix:
if we consider the pentagonal equation (\ref{pentagone}) applied to the LFT fusion matrix in the special case where
$\al_2=-\frac{1}{2b}$, and if we use the fusion rules of \cite{FF}, then the LFT fusion matrices in  (\ref{pentagone}) 
 which contain $\al_2=-\frac{1}{2b}$ are some residues of the general fusion coefficients
 (\ref{fusionmatrix}). Their expressions are given in the Appendix A.  We then obtain an
 equation that permits us to reexpress {\it each} LFT fusion matrix in (\ref{5point})
$$
F^{L}_{\al_{21}-\frac{1}{2b},\al_{32}}\left[
\begin{array}
[c]{cc}%
\al_{3}    &  \al_2 \\
\al_{4}    &   \al_{1}-\frac{1}{2b}%
\end{array}
\right]\quad,
\qquad
F^{L}_{\al_{21}-\frac{1}{2b},\al_{32}}\left[
\begin{array}
[c]{cc}%
\al_{3}    &  \al_2 \\
\al_{4}    &   \al_{1}+\frac{1}{2b}%
\end{array}
\right]  \nonumber \\
$$
in terms of
$$
F^{L}_{\al_{21}-\frac{1}{2b},\al_{32}-\frac{1}{2b}}\left[
\begin{array}
[c]{cc}%
\al_{3}    &  \al_2-\frac{1}{2b} \\
\al_{4}    &   \al_{1}%
\end{array}
\right]\quad
and \quad
F^{L}_{\al_{21}-\frac{1}{2b},\al_{32}-\frac{1}{2b}}\left[
\begin{array}
[c]{cc}%
\al_{3}    &  \al_2+\frac{1}{2b} \\
\al_{4}    &   \al_{1}%
\end{array}
\right]  \nonumber \\
$$
Rearranging the terms together, one gets the following representation:
\begin{eqnarray}
\lefteqn{\fush{j_1}{j_2}{j_3}{j_4}{j_{21},}{j_{32}}_{\epsilon}=} \nonumber \\
& & \frac{\pi \Gamma(2+b^{-2}+2j_{21})\Gamma(-2j_{32}-b^{-2}-1)}{\sin\pi(j_1+j_2+j_3+j_4+2+b^{-2})}\times \nonumber \\
& &
\Big(\frac{\Gamma^{-1}(2+b^{-2}+j_{21}+j_1+j_2)\Gamma^{-1}(2+b^{-2}+j_{21}+j_3+j_4)}{\Gamma(-1-b^{-2}-j_{32}-j_2-j_3)
\Gamma(-1-b^{-2}-j_{32}-j_1-j_4)}\times \nonumber \\
& & \times F^{L}_{-bj_{21},-bj_{32}}\left[
\begin{array}
[c]{cc}%
-bj_3    &  -bj_{2} \\
-bj_4    &  -bj_{1}%
\end{array}
\right]
\nonumber \\
& &
-\frac{e^{-i\pi\epsilon(j_1+j_2+j_3+j_4+2+b^{-2})}}{\Gamma(j_{21}-j_1-j_2)\Gamma(j_{21}-j_3-j_4)\Gamma(1-j_{32}+j_2+j_3)
\Gamma(1-j_{32}+j_1+j_4)} \times \nonumber \\
& & \times F^{L}_{-bj_{21},-bj_{32}}\left[
\begin{array}
[c]{cc}%
-bj_{3}+\frac{1}{2b}    &  -bj_{2}+\frac{1}{2b} \\
-bj_{4}+\frac{1}{2b}    &  -bj_{1} +\frac{1}{2b}%
\end{array}
\right] \Big) \label{fush3+bis}
\end{eqnarray}
We can use the same trick again by setting this time $\al_3=-\frac{1}{2b}$ in (\ref{pentagone}), so that we can reexpress
$$
F^{L}_{\al_{21}-\frac{1}{2b},\al_{32}}\left[
\begin{array}
[c]{cc}%
\al_{3}    &  \al_2 \\
\al_{4}    &   \al_{1}-\frac{1}{2b}%
\end{array}
\right]\quad,
\qquad
F^{L}_{\al_{21}-\frac{1}{2b},\al_{32}}\left[
\begin{array}
[c]{cc}%
\al_{3}    &  \al_2 \\
\al_{4}    &   \al_{1}+\frac{1}{2b}%
\end{array}
\right]  \nonumber \\
$$
in terms of
$$
F^{L}_{\al_{21},\al_{32}}\left[
\begin{array}
[c]{cc}%
\al_{3}    &  \al_2 \\
\al_{4}-\frac{1}{2b}    &   \al_{1}%
\end{array}
\right]\quad
and \quad
F^{L}_{\al_{21},\al_{32}}\left[
\begin{array}
[c]{cc}%
\al_{3}    &  \al_2 \\
\al_{4}+\frac{1}{2b}    &   \al_{1}%
\end{array}
\right]  \nonumber \\
$$
This gives us a third representation for the $H_3^+$ fusion matrix:
\begin{eqnarray}
\lefteqn{\fush{j_1}{j_2}{j_3}{j_4}{j_{21},}{j_{32}}_{\epsilon}=} \nonumber \\
& & \frac{\pi \Gamma(-2j_{21})\Gamma(1+2j_{32})}{\sin\pi(-j_1+j_2-j_3+j_4)}\times \nonumber \\
& &
\Big(\frac{e^{i\pi\epsilon(j_{21}+j_{32}-j_1-j_3)}}{\Gamma(-j_{21}-j_1+j_2)\Gamma(-j_{21}-j_3+j_4)\Gamma(1+j_{32}-j_2+j_3)
\Gamma(1+j_{32}+j_1-j_4)}\times \nonumber \\
& & \times F^{L}_{-bj_{21}+\frac{1}{2b},-bj_{32}+\frac{1}{2b}}\left[
\begin{array}
[c]{cc}%
-bj_3+\frac{1}{2b}    &  -bj_{2} \\
-bj_4    &  -bj_{1}+\frac{1}{2b}%
\end{array}
\right]
\nonumber \\
& &
-\frac{e^{i\pi\epsilon(j_{21}+j_{32}-j_2-j_4)}}{\Gamma(-j_{21}+j_1-j_2)\Gamma(j_{21}+j_3-j_4)\Gamma(1+j_{32}+j_2-j_3)
\Gamma(1+j_{32}-j_1+j_4)} \times \nonumber \\
& & \times F^{L}_{-bj_{21}+\frac{1}{2b},-bj_{32}+\frac{1}{2b}}\left[
\begin{array}
[c]{cc}%
-bj_{3}    &  -bj_{2}+\frac{1}{2b} \\
-bj_{4}+\frac{1}{2b}    &  -bj_{1} %
\end{array}
\right] \Big) \label{fush3+ter}
\end{eqnarray}
\underline{{\bf Consistency checks and remarks:}}
\begin{enumerate}
\item
One might wonder why on the first picture we did not also consider
the case where the fusion between $\al_{21}$ and $-\frac{1}{2b}$ gives $\al_{21}+\frac{1}{2b}$ (this
comment also applies for the last picture replacing $\al_{21}$ by $\al_{32}$). It is not difficult to see
that if we do the same reasoning starting with $\al_{21}+\frac{1}{2b}$,
we would simply have to replace $j_{21}$ by $-j_{21}-1$ in the expression for the fusion matrix.
As we consider $j_{21}$ being of the form
$-\frac{1}{2}+is$ with $s$ any real number, it is enough to consider only the case where the result
 of the fusion is $\al_{21}-\frac{1}{2b}$ (resp. $\al_{32}-\frac{1}{2b}$).
\item
The case when one of the external spins corresponds to a $\sl2$ degenerate representation, the other external spins being generic can be found in \cite{FGP}. The fusion coefficients appear as residues of the general fusion coefficients, and the monodromy becomes finite dimensional. We checked in this case that it is indeed possible to find a basis for the conformal blocks in which the fusion transformation takes a block diagonal form, reproducing thus the results of \cite{FGP}.\\
The simplest cases $j_2=1/2$, $j_2=1/2b^2$ and $j_2=-k/2$ can
be found in the appendix B.
\item
The $SU(2)$ WZNW model is obtained by substituting $b$ by $ib$ and
by giving the spins $j$ half integer or integer values in the range 
$0\leq j\leq k$, the level $k$ being an integer in this case.
One also has to substitute to Liouville theory the $(k+2,1)$
Minimal Model with central charge
$$
c=1-\frac{6(p-q)^{2}}{pq}, \qquad p=k+2\equiv b^{-2},\quad q=1.
$$
The second term of the sum in (\ref{fush3+bis}) always vanishes as the fusion rules in the
$SU(2)$ WZNW model are such that
$$
\frac{1}{\Gamma(j_{21}-j_1-j_2)}=\frac{1}{\Gamma(-n)}=0
$$
with $n$ some positive integer. Only the first term of the sum
remains. Then one uses the fusion rules $j_{21}=j_1+j_2-n$ and
$j_{32}=j_3+j_2-m$ to rewrite the prefactor in front of the
Liouville fusion matrix. This
 gives the ($\epsilon$ independant) result:
\begin{eqnarray}
\lefteqn{\fussu2{j_1}{j_2}{j_3}{j_4}{j_{21},}{j_{32}}=}\nonumber \\
&& =\frac{\Gamma(2-b^{-2}+2j_{21})\Gamma(2-b^{-2}+j_{32}+j_3+j_2)\Gamma(2-b^{-2}+j_{32}+j_1+j_4) }{\Gamma(2-b^{-2}+2j_{32})\Gamma(2-b^{-2}+j_{21}+j_1+j_2)\Gamma(2-b^{-2}+j_{21}+j_3+j_4)}\times \quad \nonumber\\
&& \times \quad F^{MM}_{-bj_{21},-bj_{32}}\left[
\begin{array}
[c]{cc}%
-bj_3   &  -bj_{2} \\
-bj_4    &  -bj_{1}%
\end{array}
\right]
\label{matricesu2}
\end{eqnarray}

\item
It is straightforward to check the following symmetry properties
\begin{eqnarray}
\fush{j_1}{j_2}{j_3}{j_4}{j_{21},}{j_{32}}_{\epsilon}=\fush{j_4}{j_3}{j_2}{j_1}{j_{21},}{j_{32}}_{\epsilon}=\fush{j_2}{j_1}{j_4}{j_3}{j_{21},}{j_{32}}_{\epsilon}
\end{eqnarray}
using the fact that these properties hold for the Liouville fusion matrix.\\

\item
Using the invariance of the LFT fusion matrix w.r.t. conformal
weights, one can notice the following additional symmetry:
\begin{eqnarray}
\fush{j_1}{j_2}{j_3}{j_4}{j_{21},}{j_{32}}_{\epsilon}=e^{-i\pi\epsilon(j_1+j_2+j_3+j_4+2+b^{-2})}\fush{\tilde{j}_1}{\tilde{j}_2}{\tilde{j}_3}{\tilde{j}_4}{j_{21},}{j_{32}}_{\epsilon}
\label{tutu}
\end{eqnarray}
where $\tilde{j}\equiv -j-\frac{k}{2}\equiv -j-1-\frac{1}{2b^{2}}$, as well as:\\
%\begin{eqnarray}
%\fush{j_1}{j_2}{j_3}{j_4}{j_{21},}{j_{32}}_{\epsilon}=e^{-i\pi\epsilon(j_{21}-j_1-j_2)}\fush{j_1}{j_2}{\tilde{j}_3}{\tilde{j}_4}{j_{21},}{\tilde{j}_{32}}_{\epsilon}
%\end{eqnarray}
%\begin{eqnarray}
%\fush{j_1}{j_2}{j_3}{j_4}{j_{21},}{j_{32}}_{\epsilon}=e^{i\pi\epsilon(j_{21}+j_{32}-j_1-j_3)}\fush{\tilde{j}_1}{j_2}{\tilde{j}_3}{j_4}{\tilde{j}_{21},}{\tilde{j}_{32}}_{\epsilon}
%\end{eqnarray}
\begin{eqnarray}
\fush{j_1}{j_2}{j_3}{j_4}{j_{21},}{j_{32}}_{\epsilon}=e^{i\pi\epsilon(j_{21}-j_1-j_2)}\fush{j_1}{j_2}{\tilde{j}_3}{\tilde{j}_4}{j_{21},}{\tilde{j}_{32}}_{-\epsilon}
\end{eqnarray}
\begin{eqnarray}
\fush{j_1}{j_2}{j_3}{j_4}{j_{21},}{j_{32}}_{\epsilon}=e^{i\pi\epsilon(j_{32}-j_2-j_3)}\fush{\tilde{j}_1}{j_2}{j_3}{\tilde{j}_4}{\tilde{j}_{21},}{j_{32}}_{-\epsilon}
\end{eqnarray}
These relations are known to exist in the $SU(2)$ WZWN model \cite{P} (see also
\cite{SB}). The difference is that in the
$SU(2)$ case, $e^{i\pi \epsilon(...)}$ is replaced by $(-1)^{(...)}$:
the reason for this is that there is no $\epsilon$ dependence in
the $SU(2)$ case, as we saw above.

\item
If $Re(j_i), i=1\dots 4$ are close enough to $-\frac{1}{2}$, one
is still in the range where both the Liouville fusion matrix that
appear in the formulae (\ref{fush3+},\ref{fush3+bis},\ref{fush3+ter}) remain holomorphic.
%\begin{eqnarray}
%\lefteqn{\fussu2{j_1}{j_2}{j_3}{j_4}{j_{21},}{j_{32}}=}\nonumber \\
%&& = \frac{\pi \Gamma(2+b^{-2}+2j_{21})\Gamma(-1-b^{-2}-2j_{32})}{\sin\pi (j_1+j_2+j_3+j_4+b^{-2})}\times \quad \frac{1}{\Gamma(2+b^{-2}+j_{21}+j_1+j_2)\Gamma(2+b^{-2}+j_{21}+j_3+j_4)}\times \quad \nonumber\\
%&& = \frac{1}{\Gamma(-j_{32}-j_2-j_3-1-b^{-2})\Gamma(-j_{32}-j_1-j_4-1-b^{-2})}\times \quad \nonumber \\
%&& \times \quad F^{MM}_{-bj_{21},-bj_{32}}\left[
%\begin{array}
%[c]{cc}%
%-bj_3   &  -bj_{2} \\
%-bj_4    &  -bj_{1}%
%\end{array}
%\right]
%\nonumber \\
%\end{eqnarray}

\item
Hexagonal equation:\\
We believe that the Moore-Seiberg equations continue to hold even in non-compact conformal field theories, for coherency of the operator algebra (there is no proof for this statement in general). In the special case of LFT, it was proven in \cite{PT1, PT2}; this allows us to derive the following equation for the monodromy of a 5 point
conformal block in LFT:
\begin{eqnarray}
\lefteqn{\mon{\al_1}{\al_2}{\al_4}{\al_3}{\al_{21},}{\beta}_{\epsilon}e^{i\pi\epsilon(\de_{\al_1}+\de_{\al_2}+\de_{\al_3}+\de_{\al_4}-\de_{\al_{21}-\frac{1}{2b}}-\de_{\beta})}
=}\nonumber \\
&& \int_{\frac{Q}{2}-i\infty}^{\frac{Q}{2}+i\infty}d\al_{32}\mon{\al_1}{\al_2}{\al_3}{\al_4}{\al_{21},}{\al_{32}}_{\epsilon}\mon{\al_1}{\al_4}{\al_2}{\al_3}{\al_{32}-\frac{1}{2b},}{\beta}_{\epsilon}e^{i\pi \epsilon\de_{\al_{32}-\frac{1}{2b}}}
\end{eqnarray}
As the monodromy for a 5 point function in LFT depends on braiding ({\it i.e.} depends on $\epsilon$),
it fixes the $\epsilon$ appearing in the exponential, which should be the same as the one parametrizing the LFT monodromy.\\
It is possible to derive the relation
\begin{eqnarray}
\mon{\al_1}{\al_4}{\al_2}{\al_3}{\al_{32}-\frac{1}{2b},}{\beta}_{\epsilon}=
\mon{\al_3}{\al_2}{\al_4}{\al_1}{\al_{32},}{\beta}_{\epsilon}e^{i\pi\epsilon
b^{-1}\beta} \label{lulu}
\end{eqnarray}
this leads us to the hexagonal equation for $\fush{j_1}{j_2}{j_3}{j_4}{j_{21},}{q}_{\epsilon}$
\begin{eqnarray}
\lefteqn{\fush{j_1}{j_2}{j_4}{j_3}{j_{21},}{q}_{\epsilon}e^{i\pi\epsilon(\Delta_{j_1}+\Delta_{j_2}+\Delta_{j_3}+\Delta_{j_4}-\Delta_{j_{21}}-\Delta_{q}-j_1-j_2-j_3-j_4+j_{21}+q)}=} \nonumber \\
&& \int_{-\frac{1}{2}-i\infty}^{-\frac{1}{2}+i\infty}dj_{32}\fush{j_1}{j_2}{j_3}{j_4}{j_{21},}{j_{32}}_{\epsilon}\fush{j_1}{j_4}{j_2}{j_3}{j_{32},}{q}_{\epsilon}e^{i\pi \epsilon(\Delta_{j_{32}}-j_{32})}
\end{eqnarray}
where $\beta$ and $q$ are related by $\beta=-bq+\frac{1}{2b}$.\\
Let us note that the situation here seems to be different from the already
 known cases (rational CFT's and LFT):
appearently, the fusion matrix, for $\epsilon$ given, satisfies one hexagonal
 equation {\it only}, and not two. But as $\epsilon$ can take two values +1 and -1,
 we indeed have two hexagonal equations in this model.

\item
Pentagonal equation:\\
I don't have much to say about it as I don't know how to prove it. It has to hold, as I believe the theory would be dead otherwise.

\item
Let us introduce for short $\delta=j_1+j_2+j_3+j_4+2+b^{-2}$, and let us consider for example (\ref{tutu}). This equation is a consequence of the following equality between conformal blocks:
\begin{eqnarray}
\lefteqn{\mathcal{G}^{s}_{j_{21}}(j_1,j_2,j_3,j_4|x,z)=
(xz^{-1}-1)^{\delta}x^{2(j_1+j_2+1+\frac{1}{2b^2})}
(1-x)^{2(j_3+j_2+1+\frac{1}{2b^2})}}\nonumber \\
 && \times \quad z^{-(j_1+j_2+1+\frac{1}{2b^2})}(1-z)^{-(j_3+j_2+1+\frac{1}{2b^2})}
\quad
\mathcal{G}^{s}_{j_{21}}(\tilde{j}_1,\tilde{j}_2,\tilde{j}_3,\tilde{j}_4|x,z)
\label{interesting}
\end{eqnarray}
We now insert this expression into a four point function
\begin{eqnarray}
\lefteqn{\Phi_{j_4,j_3,j_2,j_1}(x,\bar{x}|z,\bar{z})=}\nonumber \\
&& =\int_{-\frac{1}{2}+i\mathbb{R}}dj_{21}B(j_{21})C^{H_3^{+}}(j_4,j_3,j_{21})C^{H_3^{+}}(j_{21},j_2,j_1)|
\mathcal{G}^{s}_{j_{21}}(j_1,j_2,j_3,j_4|x,z)|^{2}\nonumber \\
\end{eqnarray}
We can rewrite this equation thanks to the following property of
the three point function
\begin{eqnarray}
C^{H_3^{+}}(j_3,\tilde{j}_2,\tilde{j}_1)=(\nu(b))^{-b^{-2}}B(j_2)B(j_1)C^{H_3^{+}}(j_3,j_2,j_1)
\end{eqnarray}
where $B(j)$ is the two point function.\\
It follows
\begin{eqnarray}
\lefteqn{\Phi_{j_4,j_3,j_2,j_1}(x,\bar{x}|z,\bar{z}) =\left(B(j_4)B(j_3)B(j_2)B(j_1)\right)^{-1}|z|^{-2(j_1+j_2+1+\frac{1}{2b^2})}
|1-z|^{-2(j_3+j_2+1+\frac{1}{2b^2})}}\nonumber \\
&&\times |xz^{-1}-1|^{2\delta}|x|^{4(j_1+j_2+1+\frac{1}{2b^2})}
|1-x|^{4(j_3+j_2+1+\frac{1}{2b^2})}(\nu(b))^{2b^{-2}}\times\nonumber \\
 &&
\int_{-\frac{1}{2}+i\mathbb{R}}dj_{21}B(j_{21})
C^{H_3^{+}}(\tilde{j}_4,\tilde{j}_3,j_{21})C^{H_3^{+}}(j_{21},\tilde{j}_2,\tilde{j}_1)|
\mathcal{G}^{s}_{j_{21}}(\tilde{j}_1,\tilde{j}_2,\tilde{j}_3,\tilde{j}_4|x,z)|^{2}\nonumber \\
\end{eqnarray}
from which we conclude the following relation holding at the level
of correlation functions between
$\Phi_{j_4,j_3,j_2,j_1}(x,\bar{x}|z,\bar{z})$ and
$\Phi_{\tilde{j}_4,\tilde{j}_3,\tilde{j}_2,\tilde{j}_1}(x,\bar{x}|z,\bar{z})$:
\begin{eqnarray}
\lefteqn{\Phi_{j_4,j_3,j_2,j_1}(x,\bar{x}|z,\bar{z})
=\left(B(j_4)B(j_3)B(j_2)B(j_1)\right)^{-1}
\Phi_{\tilde{j}_4,\tilde{j}_3,\tilde{j}_2,\tilde{j}_1}(x,\bar{x}|z,\bar{z})}\nonumber \\
 && \times |xz^{-1}-1|^{2\delta}
|x|^{4(j_1+j_2+1+\frac{1}{2b^2})} |1-x|^{4(j_3+j_2+1+\frac{1}{2b^2})}(\nu(b))^{2b^{-2}}\nonumber \\
&& \times |z|^{-2(j_1+j_2+1+\frac{1}{2b^2})}
|1-z|^{-2(j_3+j_2+1+\frac{1}{2b^2})}.\nonumber \\
 \label{truc}
\end{eqnarray}

%\item
%The equation (\ref{lulu}) also leads to an interesting indentity
%for the four point function (to be checked again, likely slightly
%uncorrect)
%\begin{eqnarray}
%\lefteqn{\Phi_{j_4,j_3,j_2,j_1}(x,z) =}\nonumber \\
%&& =\int_{-\frac{1}{2}+i\mathbb{R}}dj_{21}B(j_{21})C^{H_3^{+}}(j_4,j_3,j_{21})C^{H_3^{+}}(j_{21},j_2,j_1)|
%\mathcal{G}^{s}_{j_{21}}(j_1,j_2,j_3,j_4|x,z)|^{2}\nonumber \\
%&& =|x|^{4j_2}|1-x|^{4j_2}|z|^{-2j_2}|1-z|^{-2j_2}\times \quad \left(B(j_4)B(j_1)\right)^{-1}\times \quad \nonumber \\
%&& \times \quad \int_{-\frac{1}{2}+i\mathbb{R}}dj_{21}B(\tilde{j}_{21})C^{H_3^{+}}(\tilde{j}_4,j_3,\tilde{j}_{21})
%C^{H_3^{+}}(\tilde{j}_{21},j_2,\tilde{j}_1)|\mathcal{G}^{s}_{\tilde{j}_{21}}(\tilde{j}_1,j_2,j_3,\tilde{j}_4|\frac{z}{x},z)|^{2}\nonumber \\
%\end{eqnarray}
%This gives us
%\begin{eqnarray}
%\Phi_{j_4,j_3,j_2,j_1}(x,z)= |x|^{4j_2}|1-x|^{4j_2}|z|^{-2j_2}|1-z|^{-2j_2}\left(B(j_4)B(j_1)\right)^{-1}\Phi_{\tilde{j}_4,j_3,j_2,\tilde{j}_1}(zx,z)\nonumber \\
%\end{eqnarray}
\end{enumerate}

\section{Study of the singular behavior $x \rightarrow z$}
Let us denote $\psi_{\al_1,\al_{21}}^{\al_2}(z)$ the Liouville chiral vertex operators. They satisfy the operator product expansion:
\begin{eqnarray}
\psi_{\al_1,\al_{21}}^{\al_2}(z)\psi_{\al_{21},\al_{21}-\frac{1}{2b}}^{-\frac{1}{2b}}(x)\sim_{x\to z} (x-z)^{b^{-1}\al_2}\fus{\al_{21}-\frac{1}{2b}}{-\frac{1}{2b}}{\al_2}{\al_1}{\al_{21},}{\al_{2}-\frac{1}{2b}}\psi_{\al_1,\al_{21}-\frac{1}{2b}}^{\al_2-\frac{1}{2b}}(z)+\nonumber \\
+ \quad
(x-z)^{\frac{1}{b}(Q-\al_2)}\fus{\al_{21}-\frac{1}{2b}}{-\frac{1}{2b}}{\al_2}{\al_1}{\al_{21},}{\al_{2}+\frac{1}{2b}}\psi_{\al_1,\al_{21}-\frac{1}{2b}}^{\al_2+\frac{1}{2b}}(z)
\end{eqnarray}
If we insert this relation into (\ref{corres}),
we find that the asymptotic behavior when
$x\rightarrow z$ of the 4 point conformal block in the $H_3^{+}$
WZNW is related to the 4 point conformal block in LFT:
\begin{eqnarray}
\lefteqn{\mathcal{G}^{s}_{j_{21}}(j_1,j_2,j_3,j_4|x,z)\sim_{x\to z}
%(x-z)^{-b^{-1}\al_2}z^{-\frac{1}{2}\gamma_{12}-b^{-1}\al_1}(1-z)^{-\frac{1}{2}\gamma_{13}-b^{-1}\al_3}\quad\times}
%\nonumber \\
%&& \times \quad\Bigg[(x-z)^{b^{-1}\al_2}\fus{\al_{21}-\frac{1}{2b}}{-\frac{1}{2b}}{\al_2}{\al_1}{\al_{21},}{\al_{2}-
%\frac{1}{2b}}\mathcal{F}^{s}_{\al_{21}-\frac{1}{2b}}(\al_1,\al_2-\frac{1}{2b},\al_3,\al_4|z)\nonumber \\
%&&+ \quad (x-z)^{\frac{1}{b}(-\al_2+Q)}\fus{\al_{21}-\frac{1}{2b}}{-\frac{1}{2b}}{\al_2}{\al_1}{\al_{21},}{\al_{2}+
%\frac{1}{2b}}\mathcal{F}^{s}_{\al_{21}-\frac{1}{2b}}(\al_1,\al_2+\frac{1}{2b},\al_3,\al_4|z)\Bigg]\nonumber \\
%\end{eqnarray}
%\begin{eqnarray}
  z^{-\frac{1}{2}\gamma_{12}-b^{-1}\al_1}(1-z)^{-\frac{1}{2}\gamma_{13}-b^{-1}\al_3}\times }\nonumber \\
&& \times
\quad\Bigg[\fus{\al_{21}-\frac{1}{2b}}{-\frac{1}{2b}}{\al_2}{\al_1}{\al_{21},}{\al_{2}-\frac{1}{2b}}
\mathcal{F}^{s}_{\al_{21}-\frac{1}{2b}}(\al_1,\al_2-\frac{1}{2b},\al_3,\al_4|z)\nonumber \\
&&+\quad (x-z)^{\frac{1}{b}(-2\al_2+Q)}\fus{\al_{21}-\frac{1}{2b}}{-\frac{1}{2b}}{\al_2}{\al_1}{\al_{21},}{\al_{2}+\frac{1}{2b}}\mathcal{F}^{s}_{\al_{21}-\frac{1}{2b}}(\al_1,\al_2+\frac{1}{2b},\al_3,\al_4|z)\Bigg]\nonumber \\
\label{lala}
\end{eqnarray}
The first term of the sum is regular at $x=
z$, whereas the second one is singular.\\
It is easy to see that the Liouville conformal block $\mathcal{F}^{s}_{\al_{21}-\frac{1}{2b}}(\al_1,\al_2-\frac{1}{2b},\al_3,\al_4|z)$ has the same monodromy as $\mathcal{F}^{s}_{-bj_{21}}(-bj_1,-bj_2,-bj_3,-bj_4|z)$; if we then study
 the behavior when $z
\rightarrow 0$ of the first Liouville conformal block multiplied
by the spatial factor in front of the bracket, we then see that we have indeed the equality
\begin{eqnarray}
\lefteqn{\mathcal{F}^{s}_{-bj_{21}}(-bj_1,-bj_2,-bj_3,-bj_4|z)=}\nonumber\\
&&
z^{-\frac{1}{2}\gamma_{12}-b^{-1}\al_1}(1-z)^{-\frac{1}{2}\gamma_{13}-b^{-1}\al_3}
 \mathcal{F}^{s}_{\al_{21}-\frac{1}{2b}}(\al_1,\al_2-\frac{1}{2b},\al_3,\al_4|z)
\end{eqnarray}
A similar
study for the second Liouville conformal block of the sum allows us to rewrite
(\ref{lala}) as
\begin{eqnarray}
\lefteqn{\mathcal{G}^{s}_{j_{21}}(j_1,j_2,j_3,j_4|x,z)}\nonumber \\
&& \sim_{x\to z}\Bigg[\frac{\Gamma(2+b^{-2}+2j_{21})\Gamma(2+b^{-2}+j_{1}+j_2+j_3+j_4)}{\Gamma(2+b^{-2}+j_{21}+j_1+j_2)
\Gamma(2+b^{-2}+j_{21}+j_3+j_4)}\mathcal{F}^{s}_{-bj_{21}}(-bj_1,-bj_2,-bj_3,-bj_4|z)+\nonumber \\
&& +\frac{\Gamma(2+b^{-2}+2j_{21})\Gamma(-2-b^{-2}-j_{1}-j_2-j_3-j_4)}{\Gamma(j_{21}-j_1-j_2)\Gamma(j_{21}-j_3-j_4)}
 \mathcal{F}^{s}_{-bj_{21}}(-b\tilde{j}_1,-b\tilde{j}_2,-b\tilde{j}_3,-b\tilde{j}_4|z)\times \quad\nonumber \\
&&\times \quad
(x-z)^{\delta}z^{-\delta+j_1+j_2+1+\frac{1}{2b^2}}(1-z)^{j_3+j_2+1+\frac{1}{2b^2}}\Bigg]
\label{titi}
\end{eqnarray}

\underline{{\bf Remarks:}}
\begin{enumerate}
\item
It is straightforward to check the property
\begin{eqnarray}
\lefteqn{\mathcal{G}^{s}_{j_{21}}(\tilde{j}_1,\tilde{j}_2,\tilde{j}_3,\tilde{j}_4|x,z)\sim_{x\to z}}\nonumber\\
&&
(x-z)^{-\delta}z^{\delta-(j_1+j_2+1+\frac{1}{2b^2})}(1-z)^{-(j_2+j_3+1+\frac{1}{2b^2})}
\mathcal{G}^{s}_{j_{21}}(j_1,j_2,j_3,j_4|x,z) \nonumber \\
\label{trac}
\end{eqnarray}
We recover here a straighforward consequence of equation (\ref{interesting}).
\item
Let us consider again the case of the  $SU(2)$ WZNW model: for the
same reason we mentionned previously, the factor
$$\frac{1}{\Gamma(j_{21}-j_1-j_2)}$$ makes the singular term of
(\ref{titi}) vanish. Then one sees that up to a normalization of
the chiral vertex operators preserving the polynomial equations,
$\mathcal{G}(z)$ is nothing but the conformal block
$$\mathcal{F}^{s}_{-bj_{21}}(-bj_1,-bj_2,-bj_3,-bj_4|z)$$ of the $(k+2,1)$ minimal
model.
\item
Degenerate representations and fusion rules for $\sl2$ are well known \cite{AY}; in appendix C we show how this relation permits us to recover them.
\item
It is instructive to use relation (\ref{titi}) to express the
behavior when $x \to z$ of a correlation function in $H_3^{+}$ in
terms of {\it two} correlation functions in Liouville field
theory.\\
We consider
\begin{eqnarray}
\lefteqn{\Phi_{j_4,j_3,j_2,j_1}(x,\bar{x}|z,\bar{z})}\nonumber \\
&& =\int_{\mathcal{D}}dj_{21}B(j_{21})C^{H_3^{+}}(j_4,j_3,j_{21})C^{H_3^{+}}(j_{21},j_2,j_1)|
\mathcal{G}^{s}_{j_{21}}(j_1,j_2,j_3,j_4|x,z)|^{2},\nonumber \\
\end{eqnarray}
the external spins are such that $Re(j_i), i=1,\dots 4$ are close enough to $-\frac{1}{2}$,
 and $\mathcal{D}=-\frac{1}{2}+i\mathbb{R}$.
%We will take the external spins {\it real} and in the range
%$$
%\begin{array}{ll}
%-\frac{3}{2}-\frac{1}{2b^2}<j_1+j_2<-\frac{1}{2}-\frac{1}{2b^2},\qquad   &|j_1-j_2|<\frac{1}{2},\\
% -\frac{3}{2}-\frac{1}{2b^2}<j_3+j_4<-\frac{1}{2}-\frac{1}{2b^2},\qquad    &|j_3-j_4|<\frac{1}{2}.\\
%\end{array}
%$$
Inserting (\ref{titi}), one then sees that the prefactor in front the
conformal block\\
$|\mathcal{F}^{s}_{-bj_{21}}(-bj_1,-bj_2,-bj_3,-bj_4|z)|^{2}$
multiplied by the $H_3^{+}$ three point function recombines to
give the Liouville three point function
$$C^{L}(-bj_{4},-bj_3,Q+bj_{21})
C^{L}(-bj_{21},-bj_2,-bj_1).$$  Similarly the prefactor of the
conformal block
$|\mathcal{F}^{s}_{-bj_{21}}(-b\tilde{j}_1,-b\tilde{j}_2,-b\tilde{j}_3,-b\tilde{j}_4|z)|^{2}
$ (using the property that the Liouville conformal blocks depend on the
Liouville conformal weights) multiplied by the $H_3^{+}$ three
point function recombines to give the Liouville three point
function
$$\left(B(j_1)B(j_2)B(j_3)B(j_4)\right)^{-1}C^{L}(-b\tilde{j}_{4},-b\tilde{j}_3,Q+bj_{21})
C^{L}(-bj_{21},-b\tilde{j}_2,-b\tilde{j}_1)$$
We have
\begin{eqnarray}
\lefteqn{\Phi_{j_4,j_3,j_2,j_1}(x,\bar{x}|z,\bar{z})\sim_{x\to z}}\nonumber \\
&&
a\int_{\mathcal{D}}dj_{21}C^{L}(-bj_{4},-bj_3,Q+bj_{21})C^{L}(-bj_{21},-bj_2,-bj_1)|
\mathcal{F}^{s}_{-bj_{21}}(z)|^{2}+\nonumber \\
&& + \quad |x-z|^{2\delta}|z|^{-2\delta +
2(j_1+j_2+1+\frac{1}{2b^2})}
|1-z|^{2(j_3+j_2+1+\frac{1}{2b^2})}\times \quad \left(B(j_1)B(j_2)B(j_3)B(j_4)\right)^{-1}\times \quad \nonumber \\
&& \times \quad b\int_{\mathcal{D}}dj_{21}
C^{L}(-b\tilde{j}_{4},-b\tilde{j}_3,Q+bj_{21})C^{L}(-bj_{21},-b\tilde{j}_2,-b\tilde{j}_1)|\tilde{\mathcal{F}}^{s}_{-bj_{21}}(z)|^{2}\nonumber \\
\end{eqnarray}
with
$a=\gamma(j_1+j_2+j_3+j_4+2+b^{-2}),\quad
b=\gamma(-j_1-j_2-j_3-j_4-2-b^{-2})$.
A LFT four point correlation function should
factorize over the domain
$\mathcal{D'}=-\frac{1}{2}-\frac{1}{2b^{2}}+i\mathbb{R}$; so we have to
deform the
contour of integration from $\mathcal{D}=-\frac{1}{2}+i\mathbb{R}$
to $\mathcal{D'}=-\frac{1}{2}-\frac{1}{2b^{2}}+i\mathbb{R}$ in order
to rewrite the expression in terms of correlation functions in LFT. While
we
deform the contour, we pick up a finite number of those
poles $-bj_p$ which
 are in the range $\frac{b}{2}<Re(-bj_p)<\frac{Q}{2}$, that come
 from the Liouville three point functions of the regular term (there are
no poles coming from the Liouville three point functions of the
singular term in this case). It would be of course possible to
give whatever values we like for the external spins (for example
we could consider them to be real), then the residues that would
be picked up depend on
 a case by case study, as the poles $j_p$ depend on the values of
 the external spins $j_1\dots j_4$.

%\begin{eqnarray}
%-bj_{21}&=&Q+bj_1+bj_2-nb \nonumber\\
%-bj_{21}&=&Q+bj_3+bj_4-mb,
%\end{eqnarray}
%where $n,m$ are some positive integers. There are no poles coming
%from
 %$$\left(B(j_1)B(j_2)B(j_3)B(j_4)\right)^{-1}C^{L}(-b\tilde{j}_{4},-b\tilde{j}_3,Q+bj_{21})
 %C^{L}(-bj_{21},-b\tilde{j}_2,-b\tilde{j}_1).$$\\

Hence we can rewrite the behavior of the 4 point function in the $H_3^+$ WZNW
model in terms of 4 point functions in LFT:
\begin{eqnarray}
\lefteqn{\Phi_{j_4,j_3,j_2,j_1}(x,\bar{x}|z,\bar{z})
\sim_{x\to z} a\mathcal{V}_{-bj_4,-bj_3,-bj_2,-bj_1}(z,\bar{z})+
b\mathcal{V}_{-b\tilde{j}_4,-b\tilde{j}_3,-b\tilde{j}_2,-b\tilde{j}_1}(z,\bar{z})\times}\nonumber\\
 && \left(B(j_1)B(j_2)B(j_3)B(j_4)\right)^{-1}\times  |x-z|^{2\delta}|z|^{-2\delta+
2(j_1+j_2+1+\frac{1}{2b^2})} |1-z|^{2(j_3+j_2+1+\frac{1}{2b^2})}\nonumber \\
&&+ (\mathrm{Residues}).
\end{eqnarray}

\item
If we make the same analysis as above studying this time the
behavior when $x \rightarrow 1$ of the conformal block, we find
(this is a consequence of a relation already noticed in
\cite{PS}):
\begin{eqnarray}
\mathcal{G}^{s}_{j_{21}}(j_1,j_2,j_3,j_4|x,z)\sim_{x\to 1}
z^{-j_2}(1-z)^{-j_2}\mathcal{G}^{s}_{\tilde{j}_{21}}(\tilde{j}_1,j_2,j_3,\tilde{j}_4|\frac{z}{x},z)
\end{eqnarray}
It is of course possible to make an analysis similar to the one made above to get some relation at the level of correlation functions.
\end{enumerate}

\section*{Concluding remarks}
There are several points that deserve further study:
\begin{itemize}
\item
It remains of course to understand precisely the physical meaning of 
the singularity of the conformal blocks when $x \sim z$.
\item
It is straightforward to construct in the case of the spherical branes 
\cite{PST}
the boundary three point function along the lines of 
 \cite{PT3}: the normalization for the boundary operators was computed in \cite{PST},
 and the fusion matrix in this case is like
 the one of equation (\ref{matricesu2}) (by changing $b^{-2}$ into $-b^{-2}$, and the $(k+2,1)$ minimal model replaced by LFT). 
 As for the $AdS_2$ branes, it seems to be a problem to construct
 a cyclic invariant boundary three point function, as well as to
  recover the boundary two
 point function of \cite{PST}.
 \item
It is worth trying to generalize this method to the supersymmetric case:
 this could maybe lead to a proof
 at the level of correlation functions for the duality between
N=2 Liouville and the superconformal $SL(2)/U(1)$ model
\cite{GKP}; maybe the results presented here work can also help at
proving rigorously the duality conjectured by \cite{FZZ2} between
the sin-Liouville theory and $SL(2)/U(1)$ WZNW model.
\item
It would be very interesting to build the fusion matrix as a
 6j symbol of a quantum group. The proposal of \cite{FM} is
 the "pair" $U_q(sl(2)),U_{q'}(osp(1|2))$. In particular, such a construction 
 would ensure the validity of the pentagonal equation. We hope to be able
to say more about this problem in the future.
\end{itemize}

\section*{Acknowledgments}
It is a pleasure to thank V.~A.~Fateev, I.~Kostov, G.~Mennessier, V.~Schomerus, Ch.~Schweigert, Al.~B.~Zamolodchikov and J.-B.~Zuber
for discussions and interest in
this work. \\
The author would like to thank V.~B.~Petkova for clarifying correspondence and for pointing out useful references after the 
first version of this paper had appeared.\\
Work supported by EU under contract HPRN-CT-2000-00122.

\newpage
\section*{Appendix A. Some residues of the Liouville fusion matrix}

It is well known that in the case where one of $\al_1, \dots,\al_4$, say $\al_i$ equals $-\frac{n}{2}b-\frac{m}{2}b^{-1}$ where $n,m \in \mathbb{N}$ and where
a triple $(\Delta_{\al_4},\Delta_{\al_3},\Delta_{\al_{21}})$, $(\Delta_{\al_{21}},\Delta_{\al_2},\Delta_{\al_{1}})$ which contains $\Delta_{\al_i}$ satisfies the fusion rules of \cite{FF}, one will find that the fusion coefficients that multiply the conformal blocks are residues of the general fusion coefficient.\\
In the case where $\al_2=-\frac{1}{2}b$, the fusion rules are:
$$\left\lbrace
\begin{array}{l}
\alpha_{21} = \alpha_{1} -s\frac{b}{2}\\
\alpha_{32} = \alpha_{3} -s'\frac{b}{2} \\
\end{array}\right.$$
where $s,s'=\pm$.\\
There are four entries for the fusion matrix in this special case
$$
\fus{\al_1}{-b/2}{\al_3}{\al_4}{\al_1-sb/2,}{\al_3-s'b/2} \equiv F^{L}_{s,s'}
$$
which expressions are well known to be:
\begin{eqnarray}
F_{++}&=&\frac{\Gamma(b(2\al_1-b))\Gamma(b(b-2\al_3)+1)}{\Gamma(b(\al_1-\al_3-\al_4+b/2)+1)\Gamma(b(\al_1-\al_3+\al_4-b/2))} \nonumber \\
F_{+-}&=& \frac{\Gamma(b(2\al_1-b))\Gamma(b(2\al_3-b)-1)}{\Gamma(b(\al_1+\al_3+\al_4-3b/2)-1)\Gamma(b(\al_1+\al_3-\al_4-b/2))} \nonumber \\
F_{-+}&=&\frac{\Gamma(2-b(2\al_1-b))\Gamma(b(b-2\al_3)+1)}{\Gamma(2-b(\al_1+\al_3+\al_4-3b/2))\Gamma(1-b(\al_1+\al_3-\al_4-b/2))} \nonumber \\
F_{--}&=& \frac{\Gamma(2-b(2\al_1-b))\Gamma(b(2\al_3-b)-1)}{\Gamma(b(-\al_1+\al_3+\al_4-b/2))\Gamma(b(-\al_1+\al_3-\al_4+b/2)+1)} \nonumber \\
\label{f1/2}
\end{eqnarray}
The dual case where $\al_2=-b^{-1}/2$ is obtained by substituting $b$ by $b^{-1}$.

\section*{Appendix B. Special cases of the $H_3^{+}$ fusion matrix}
Degenerate representations and fusion rules are well known for $\hat{sl}(2)$ \cite{AY}. We will
study three easy cases.
\begin{enumerate}
\item $j_2=1/2$\\
This case was first derived in \cite{FZ}.\\
The fusion rules are $ j_{21}=j_1+1/2\quad (+),\quad
j_{21}=j_1-1/2\quad (-). $ We are in the case where the second term of the sum (\ref{fush3+bis}) always vanishes. It is
straightforward to use the results of Appendix A to get
\begin{eqnarray}
F^{H_3^+}_{++}&=&\frac{\Gamma(b^2(-2j_1-1))\Gamma(b^2(1+2j_3)+1)}
{\Gamma(b^2(-j_1+j_3+j_4+1/2)+1)\Gamma(b^2(-j_1+j_3-j_4-1/2))} \nonumber \\
F^{H_3^+}_{+-}&=& \frac{\Gamma(b^2(-2j_1-1))\Gamma(b^2(-2j_3-1))}
{\Gamma(-b^2(j_1+j_3+j_4+3/2))\Gamma(b^2(-j_1-j_3+j_4-1/2)} \nonumber \\
F^{H_3^+}_{-+}&=&\frac{\Gamma(1+b^2(1+2j_1))\Gamma(1+b^2(1+2j_3))}
{\Gamma(1+b^2(j_1+j_3+j_4+3/2)))\Gamma(1+b^2(j_1+j_3-j_4+1/2))} \nonumber \\
F^{H_3^+}_{--}&=& -\frac{\Gamma(1+b^2(1+2j_1))\Gamma(-b^2(2j_3+1))}
{\Gamma(b^2(j_1-j_3-j_4-1/2))\Gamma(b^2(j_1-j_3+j_4+1/2)+1)} \nonumber \\
\nonumber
\end{eqnarray}
\item $j_2=1/2b^2$\\
This case can be found in \cite{FGP}, but as we use the same normalisation as 
\cite{T2}, we shall compare directly with this last paper.\\
The fusion rules are
$$
j_{21}=j_1+\frac{1}{2b^2} \quad(+),\quad
j_{21}=j_1-\frac{1}{2b^2}\quad (-),\quad j_{21}=-j_1-k/2\quad
(\mathrm{x}).
$$
We find it more convenient to work directly with the
representation (\ref{fush3+bis}). The second LFT fusion matrix of
the sum is equal to 1 in the cases:
\begin{eqnarray}
j_{21}=j_1-\frac{1}{2b^2},\quad
\mathrm{or}\quad j_{21}=-j_1-1-\frac{1}{2b^2}\nonumber \\
j_{32}=j_3-\frac{1}{2b^2},\quad \mathrm{or}\quad
j_{32}=-j_3-1-\frac{1}{2b^2},\nonumber
\end{eqnarray}
and equal to zero otherwise. It is then straightforward to compute
\begin{eqnarray}
F^{H_3^+}_{++}&=&\frac{\Gamma(-2j_1-\b)\Gamma(2j_3+\b+1)}
{\Gamma(-j_1+j_3+j_4+\b/2+1)\Gamma(-j_1+j_3-j_4-\b/2)} \nonumber \\
F^{H_3^+}_{+-}&=&\frac{\Gamma(-2j_1-\b)\Gamma(-2j_3-1)}
{\Gamma(-j_1-j_3-j_4-1-\b/2)\Gamma(-j_1-j_3+j_4-\b/2)} \nonumber \\
F^{H_3^+}_{+\mathrm{x}}&=&\frac{\Gamma(-2j_1-\b)\Gamma(2j_3+1)\Gamma(-2j_3-1-\b)}
{\Gamma(-\b)\Gamma(-j_1+j_3-j_4-\b/2)\Gamma(-j_1-j_3+j_4-\b/2)} \nonumber \\
F^{H_3^+}_{-+}&=&\frac{\Gamma(2j_1+2)\Gamma(2j_3+\b+1)}
{\Gamma(j_1+j_3+j_4+\b/2+2)\Gamma(j_1+j_3-j_4+1+\b/2)} \nonumber \\
F^{H_3^+}_{\mathrm{x}+}&=&\frac{\Gamma(-2j_1)\Gamma(2j_1+2+\b)\Gamma(2j_3+\b+1)}
{\Gamma(1+\b)\Gamma(-j_1+j_3+j_4+\b/2+1)\Gamma(j_1+j_3-j_4+1+\b/2)} \nonumber \\
\end{eqnarray}
For other cases the second LFT fusion matrix contributes; we find:
\begin{eqnarray}
F^{H_3^+}_{--}&=&\frac{e^{-i\pi\epsilon\b}\Gamma(2j_1+2)\Gamma(-2j_3-1)}
{\Gamma(j_1-j_3-j_4-\b/2)\Gamma(j_1-j_3+j_4+1+\b/2)} \nonumber \\
F^{H_3^+}_{-\mathrm{x}}&=-&\frac{e^{-i\pi\epsilon(2j_3+\b)}\Gamma(2j_1+2)\Gamma(2j_3+1)\Gamma(-2j_3-1-\b)}
{\Gamma(-\b)\Gamma(j_1+j_3+j_4+2+\b/2)\Gamma(j_1-j_3-j_4-\b/2)} \nonumber \\
F^{H_3^+}_{\mathrm{x}-}&=-&\frac{e^{-i\pi\epsilon(2j_1+\b)}\Gamma(-2j_1)\Gamma(2j_1+2+\b)\Gamma(-2j_3-1)}
{\Gamma(1+\b)\Gamma(-j_1-j_3-j_4-1-\b/2)\Gamma(j_1-j_3+j_4+1+\b/2)} \nonumber \\
F^{H_3^+}_{\mathrm{x}\mathrm{x}}&=&\Gamma(-2j_1)\Gamma(2j_3+1)\Gamma(2j_1+2+\b)\Gamma(-2j_3-1-\b)\nonumber\\
&\times &
\frac{\sin\pi(-\b)\sin\pi(j_1-j_3-j_4-\b/2)\sin\pi(-j_1+j_3-j_4-\b/2)}{\pi^2\sin\pi(j_1+j_3+j_4+2+3\b/2)}\nonumber
\\
&-&
\frac{e^{-i\pi\epsilon(j_1+j_3+j_4+2+3\b/2)}\Gamma(-2j_1)\Gamma(2j_3+1)\sin\pi(j_1+j_3+j_4+2+\b/2)}{\Gamma(-2j_1-1-\b)\Gamma(2j_3+2+\b)\sin\pi(j_1+j_3+j_4+2+3\b/2)}\nonumber
\end{eqnarray}
These coefficients are in agreement with \cite{T2} for the choice
$\epsilon =1$ , excepted for $F^{H_3^+}_{\mathrm{x}\mathrm{x}}$,
where there seems to be a slight discrepancy in some of the arguments of the gamma functions. (We did not redo the computations of \cite{T2}, where the fusion matrix was obtained thanks to an explicit integral representation for the conformal blocks).
\item $j_2=-k/2$\\
This elementary case was explicitly mentionned in \cite{FZ}.
The fusion rule is $ j_{21}=-j_1-k/2. $ The first term of the sum
vanishes and the second LFT fusion matrix is equal to one. It
remains to evaluate the product of gamma functions; we find:
\begin{eqnarray}
F^{H_3^+}=e^{-i\pi \epsilon(j_1+j_3+j_4+1+\frac{1}{b^2})}
\nonumber
\end{eqnarray}
\end{enumerate}

\section*{Appendix C. Fusion rules for $\hat{sl}(2)$ algebra at generic level}
We present here a way to find degenerate representations and fusion rules for
$\sl2$ from the knowledge of the degenerate representations and fusion rules
of the Virasoro algebra.\\
We derived in the main text the following asymptotic behavior when $x\to z$ for the conformal blocks
of the $SL(2,\mathbb{C})/SU(2)$ WZNW, relating them to two Liouville conformal blocks:
\begin{eqnarray}
\lefteqn{\mathcal{G}^{s}_{j_{21}}(j_1,j_2,j_3,j_4|x,z) }\nonumber \\
&& \sim_{x\to z}\Bigg[\frac{\Gamma(2+b^{-2}+2j_{21})\Gamma(2+b^{-2}+j_{1}+j_2+j_3+j_4)}{\Gamma(2+b^{-2}+j_{21}+j_1+j_2)
\Gamma(2+b^{-2}+j_{21}+j_3+j_4)}\mathcal{F}^{s}_{-bj_{21}}(-bj_1,-bj_2,-bj_3,-bj_4|z)+\nonumber \\
&& +\frac{\Gamma(2+b^{-2}+2j_{21})\Gamma(-2-b^{-2}-j_{1}-j_2-j_3-j_4)}{\Gamma(j_{21}-j_1-j_2)\Gamma(j_{21}-j_3-j_4)}
 \mathcal{F}^{s}_{-bj_{21}}(-bJ_1,-bJ_2,-bJ_3,-bJ_4|z)\times \quad\nonumber \\
&&\times \quad
(x-z)^{j_1+j_2+j_3+j_4+2+b^{-2}}z^{-j_3-j_4-1-\frac{1}{2b^2}}(1-z)^{j_3+j_2+1+\frac{1}{2b^2}}\Bigg]
\nonumber\\
\label{trulu}
\nonumber
\end{eqnarray}
 with $J_i=j_i-\frac{1}{2b^2}$.

Although there is no closed form known for the Liouville conformal blocks, 
they are completly determined by the conformal symmetry. They depend on
 conformal weights only, {\it i.e.} are invariant when $-bj_i$ (resp. $-bJ_i$) is changed into $Q+bj_i$
(resp. $Q+bJ_i$).

\begin{enumerate}
\item
Degenerate representations of the Virasoro algebra.\\
It is well known that the case where $j_i$ (resp. $J_i$) equals
$\frac{n}{2}+\frac{m}{2}b^{-2}$
 with $n,m$ non negative integers, corresponds to a degenerate Virasoro representation $\mathcal{V}_{-bj_i}$ (resp.
  $\mathcal{V}_{-bJ_i}$).
 The conformal block $\mathcal{F}^{s}_{-bj_{21}}$ then only exists for a finite number of values of $-bj_{21}$
 \cite{FF}:
\begin{eqnarray}
-bj_{21}=-bj_1+bj_2-ub-vb^{-1}.
\label{regfus}
\nonumber
\end{eqnarray}
where $u,v$ are integers such that $0\leq u\leq n,\quad 0\leq v\leq m$.\\
It would be equivalent to write $Q+bj_{21}$ instead of $-bj_{21}$ in equation (\ref{regfus}) since the Virasoro representations $\mathcal{V}_{-bj}$ and $\mathcal{V}_{Q+bj}$ are equivalent. We will see that this fact will play an important role in the determination of the degenerate representations and fusion rules for $\sl2$.

\item
Degenerate representations and fusion rules for $\sl2$.\\
Let us call for short $F^{1}$ the first Liouville conformal block of the sum (\ref{trulu}) and the second
 one $F^{2}$.
\begin{theo}
The spin $j_{m,n}$ that labels the degenerate representation of $\sl2$ $\mathcal{P}_{j_{n,m}}$ are
also labels for the degenerate Virasoro representation $\mathcal{V}_{-bj_{n,m}}$ or
$\mathcal{V}_{-bJ_{n,m}}$.
\end{theo}
In other words, it means that $\mathcal{P}_{j_{n,m}}$ is a degenerate representation of $\sl2$ iff:
\begin{eqnarray}
(\mathrm{a})\qquad   j_{n,m} &=&\frac{n}{2}+\frac{m}{2}b^{-2}\quad \mathrm{or}\nonumber \\
(\mathrm{b})\qquad  j_{n,m} &=&-\left(\frac{n}{2}+1\right)-\left(\frac{m}{2}+\frac{1}{2}\right)b^{-2}.
\end{eqnarray}
with $n,m$ non negative integers.\\
We now provide two rules that will allow us to determine the $\sl2$ fusion rules:
\begin{montheo}
If $j_2$ is such that both $\mathcal{V}_{-bj_2}$ and $\mathcal{V}_{-bJ_2}$ correspond to degenerate
 Virasoro
 representations, then the admissible $\sl2$ fusion rules consist of the set of common fusions rules plus $j_{21}=j_1+j_2$ if $j_2$ is of the form (a), and $j_{21}=-j_1-j_2-2-b^{-2}$ if $j_2$ is
 of the form (b).
\end{montheo}
\begin{montheo}
If $j_2$ is such that $\mathcal{V}_{-bj_2}$ is a degenerate Virasoro representation
and not $\mathcal{V}_{-bJ_2}$ (or converse), then the $\sl2$ fusion rules are such that the factor
$\Gamma^{-1}(j_{21}-j_1-j_2)$ in front of $F^{2}$ should be equal to zero (resp. $\Gamma^{-1}(2+b^{-2}+j_{21}+j_1+j_2)$ in front of $F^{1}$). Note that this case happens only if m=0.
\end{montheo}
We shall start by three easy examples as the generalization is straightforward.
\begin{enumerate}
\item Examples:\\
      \begin{enumerate}
     \item $j_2=\frac{1}{2}$\\
In this case we have $-bj_2=-\frac{b}{2}$ so the Virasoro representation $\mathcal{V}_{-\frac{b}{2}}$ is
 degenerate. The fusion rules are:
\begin{eqnarray}
-bj_{21} &=& -bj_{1} -\frac{b}{2},\quad -bj_{21} = -bj_{1} +\frac{b}{2},\quad \mathrm{or}\nonumber \\
Q+bj_{21} &=& -bj_{1} -\frac{b}{2}, \quad Q+bj_{21} = -bj_{1} +\frac{b}{2}. \\
\end{eqnarray}
Let us turn to $F^{2}$: $-bJ_2=-\frac{b}{2}+\frac{1}{2b}$  does not
 correspond to any degenerate Virasoro representation.
We use rule 2 to select the admissible set of fusion rules, and find
\begin{eqnarray}
j_{21}=j_1+j_2, \quad j_{21}=j_1-j_2.
\end{eqnarray}
         \item $j_2=\frac{1}{2b^2}$\\
In this case we have $-bj_2=-\frac{1}{2b}$ so the Virasoro representation $\mathcal{V}_{-\frac{1}{2b}}$ is
 degenerate. The allowed values for $j_{21}$ are:
\begin{eqnarray}
-bj_{21} &=& -bj_{1} -\frac{1}{2b}, \quad -bj_{21} = -bj_{1} +\frac{1}{2b}, \quad \mathrm{or}
\nonumber\\
 Q+bj_{21} &=& -bj_{1} -\frac{1}{2b}, \quad Q+bj_{21} = -bj_{1} +\frac{1}{2b}.
\end{eqnarray}
As for the second term, we have $-bJ_2=0$, which corresponds to the  identity representation.
The fusion rules are  $-bj_{21}=-bj_1+\frac{1}{2b}$ or $Q+bj_{21}=-bj_1+\frac{1}{2b}$. The common set of fusion rules consists of
\begin{eqnarray}
-bj_{21} = -bj_{1} +\frac{1}{2b}, \quad -bj_{21} = bj_{1} -\frac{1}{2b}+Q .
\end{eqnarray}
According to the rule 1, we should also include
  $-bj_{21} = -bj_{1} -\frac{1}{2b}$.\\
As a conclusion, we are left with the following three possibilities:
\begin{eqnarray}
-bj_{21} = -bj_{1} -\frac{1}{2b}, \quad -bj_{21} = -bj_{1} +\frac{1}{2b}, \quad -bj_{21} = bj_{1} -
\frac{1}{2b}+Q .
\end{eqnarray}
        \item $j_2=-\frac{k}{2}$\\
In this case $-bj_2=b+\frac{1}{2b}$  does not correspond to any degenerate Virasoro representation, and
$-bJ_2=Q$. As $\mathcal{V}_Q$ and $\mathcal{V}_0$ are equivalent Virasoro representations, we see
 that $-bJ_2=Q$
actually corresponds to the identity representation. Hence the fusion rules are
$-bj_{21}=-bj_1+\frac{1}{2b}$ or $-bj_{21}=bj_1+b+\frac{1}{2b}$. The latter rule is the only
 acceptable one,
as it makes the term $\Gamma^{-1}(2+b^{-2}+j_{21}+j_1+j_2)$ in front of $F^{1}$ vanish.\\  Let us note that
the $\sl2$ representation $\mathcal{P}_{-\frac{k}{2}}$ plays a role very similar to the identity, as the
 decomposition of its tensor product with
 an arbitrary representation $\mathcal{P}_{j}$ gives the representation $\mathcal{P}_{-j-\frac{k}{2}}$ only.
        \end{enumerate}
\item General case:\\
  \begin{enumerate}
  \item $j_2=\frac{n}{2}+\frac{m}{2}b^{-2}$, with $n\in \mathbb{N}, m\in \mathbb{N}$.\\
The allowed values for $j_{21}$ are either
\begin{eqnarray}
j_{21} &=& j_1-j_2+u+vb^{-2}\quad  \mathrm{or} \nonumber \\
j_{21} &=&j_2-j_1-(u'+1)-(v'+1)b^{-2},
\end{eqnarray}
where $0\le u \le n,\quad 0\le v \le m,\quad 0\le u' \le n,\quad 0\le v' \le m-1$.
   \item $j_2=-\left(\frac{n}{2}+1\right) -\left(\frac{m}{2}+\frac{1}{2}\right)b^{-2}$,
with $n\in\mathbb{N},m\in\mathbb{N}$.\\
The allowed values for $j_{21}$ are either
\begin{eqnarray}
j_{21} &=&j_1-j_2-(U+1)-(V+1)b^{-2}\quad  \mathrm{or} \nonumber \\
j_{21}&=&j_2-j_1+U'+V'b^{-2},
\end{eqnarray}
where $0\le U \le n,\quad 0\le V \le m-1,\quad 0\le U' \le n,\quad 0\le V' \le m$.
  \end{enumerate}
These results are in agreement with \cite{AY}.
\end{enumerate}

\end{enumerate}

\end{document}